\documentclass{article}
\usepackage{graphicx}
\usepackage{spconf}
\usepackage{color}
\usepackage{float}
\usepackage{amsmath,bm,amssymb,amsfonts,amsthm,epsfig}
\usepackage{graphicx}
\usepackage{psfrag}
\usepackage[caption=false]{subfig}
\usepackage{comment}
\usepackage{url}
\usepackage{cite} 

\graphicspath{{Figures/}}
\usepackage{epstopdf}
\usepackage{lipsum}

\newtheorem*{thm*}{Theorem}

\newtheorem*{remark*}{Remark}

\usepackage[hidelinks]{hyperref}
\usepackage[capitalize]{cleveref}
\crefname{thm}{Theorem}{Theorems}

\newcounter{opteq}

\usepackage{arydshln}
\usepackage{mathrsfs}
\usepackage{bbm}

\usepackage[toc,page]{appendix}

\newcommand{\txt}[1]{\text{\normalfont #1}}

\DeclareMathOperator{\T}{\top}

\DeclareMathOperator{\tx}{t}
\DeclareMathOperator{\rx}{r}

\DeclareMathOperator{\ul}{}%ul

\DeclareMathOperator{\si}{}%si

\makeatletter
\newcounter{savesection}
\newcounter{apdxsection}
\renewcommand\appendix{\par
	\setcounter{savesection}{\value{section}}%
	\setcounter{section}{\value{apdxsection}}%
	\setcounter{subsection}{0}%
	\gdef\thesection{\@Alph\c@section}}
\newcommand\unappendix{\par
	\setcounter{apdxsection}{\value{section}}%
	\setcounter{section}{\value{savesection}}%
	\setcounter{subsection}{0}%
	\gdef\thesection{\@arabic\c@section}}
\makeatother

\usepackage[inline]{enumitem}%\let\labelindent\relax 
\usepackage{tikz}
\usepackage{pgfplots} 
\pgfplotsset{compat=1.16}
\usetikzlibrary{pgfplots.groupplots}
\pgfplotsset{
	every axis/.append style={
	label style={font=\footnotesize},
		tick label style={font=\footnotesize},  
	legend style = {font = \footnotesize},
	}
}

\usepackage[percent]{overpic}

\usepackage[hang,flushmargin]{footmisc}

\title{On array geometry and self-interference in full-duplex massive MIMO communications}
\name{\begin{tabular}{c} Robin Rajam\"{a}ki$^{\star\dagger}$
		and Risto Wichman$^\dagger$
  \end{tabular}}
\address{$^{\star}$University of California, San Diego, USA;\ $^\dagger$Aalto University, Finland}

\begin{document}
	\setlength\belowcaptionskip{-15ex}
	%	\ninept
	%
	\maketitle
	\begin{abstract}		
	This paper studies the role of the joint transmit-receive antenna array geometry in shaping the self-interference (SI) channel in full-duplex communications. We consider a simple spherical wave SI model and two prototypical linear array geometries with uniformly spaced transmit and receive antennas. We show that the resulting SI channel matrix has a regular (Toeplitz) structure in both of these cases. However, the number of significant singular values of these matrices---an indication of the severity of SI---can be markedly different. We demonstrate that both reduced SI and high angular resolution can be obtained by employing suitable sparse array configurations that fully leverage the available joint transmit-receive array aperture without suffering from angular ambiguities. Numerical electromagnetic simulations also suggest that the worst-case SI of such sparse arrays need not increase---but can actually decrease---with the number of antennas. Our findings provide preliminary insight into the extent to which the array geometry alone can mitigate SI in full-duplex massive MIMO communications systems employing a large number of antennas.
	%Self-interference is a critical problem in full-duplex (FD) systems where the continuously active transmit (Tx) and receive (Rx) antennas are near each other. It can be reduced by judiciously designing the Tx/Rx array configuration since the relative placement of the Tx and Rx antennas strongly impacts SI and, in particular, the effective rank of the SI channel matrix. A low-rank SI channel matrix enables dedicating more spatial resources toward serving users rather than unnecessarily mitigating SI. Indeed, current and emerging wireless multiantenna systems, such as massive MIMO, allow unprecedented control over the SI channel due to the large number of employed antennas. However, this also results in a vast number of possible array configurations, which presents a formidable challenge for optimizing the geometry. Consequently, this paper explores SI models that can be used as guiding principles for array design. The developed heuristics offer insight into the role of the array geometry in FD systems and can be used to predict which arrays lead to low and high SI, respectively. We also consider secondary sensing-based optimization criteria related to the Tx beam pattern and the virtual Tx-Rx array geometry and characterize the trade-off between SI and sensing performance.
	\end{abstract}
	%
%		\begin{keywords}
%		\end{keywords}
%	\setlength{\textfloatsep}{7pt}

	\section{Introduction}
    
    Full-duplex technology holds the promise of doubling the spectral efficiency of future wireless systems by utilizing the same temporal, spectral, and spatial resources for simultaneous uplink (UL) and downlink (DL) communications \cite{sabharwal2014inband}. Indeed, FD has recently experienced a surge of renewed research interest due to the ever increasing performance demands of emerging  millimeter-wave massive multiple-input multiple-output (MIMO) communications \cite{alexandropoulos2022fullduplex}, and integrated sensing and communications (ISAC) systems \cite{barneto2021fullduplex}.%, which set increasingly higher performance demands.% on the wireless system.
    
    % impeding the widespread deployment% at FD base stations and among user equipment 
    A key technical challenge of full-duplex communications is the potentially 
    severe \emph{self-interference} (SI) caused by simultaneous transmission 
    and reception at the base station (BS). Cancellation of this SI at the BS 
    receiver may further be complicated by transmitter noise and array 
    calibration errors. Moreover, canceling the SI digitally may be infeasible, 
    due to the large dynamic range that would be required of the 
    analog-to-digital converters (ADCs) at the receiver to digitize both the 
    weak received UL signal of interest and the strong transmitted DL signal 
    causing the interference. Typical approaches to SI mitigation therefore 
    include precoding at the transmitter \cite{riihonen2011mitigation},  
    analog-domain cancellation at the receiver \cite{kolodziej2019inband}, and 
    improving spatial isolation between transmit and receive antennas through
     \emph{antenna array design}.
     %RW Antenneja on suunniteltu paljonkin, mutta enemmän antennisuunnitelun 
     %näkökulmasta
     %However, there is also a third way to curb SI which remains less 
     %explored: \emph{antenna array design}. 
     Indeed, the relative placement of the transmit (Tx) and receive (Rx) 
     antennas fundamentally controls SI. 
     In \cite{choi10}, the receive antenna is placed between two transmit 
     antennas such that the two transmitted signals sum up destructively in the 
     receive antenna. The geometry does not scale up to large antenna arrays, 
     but in general,
     a favorable joint Tx-Rx array 
     geometry can reduce---potentially even eliminate---the need for additional 
     SI cancellation, reducing hardware costs and freeing up resources for 
     communications tasks \cite{everett2016softnull}. SI mitigation via array 
     design holds great promise in future 6G and beyond massive MIMO systems, 
     which offer an unprecedented number of degrees of freedom to shape the SI 
     channel due to the hundreds or even thousands of antennas envisioned to be 
     employed at a single BS \cite{larsson2014massive}. %Analog/hybrid 
     %beamforming is a prominent example of the latter.
  	%RW Jos nyt jotain yrittäisi sanoa polarisaatiosta.
  	%Spatial isolation could be further improved by using different 
  	%polarizations 
  	%in Tx and Rx arrays. However, dual-polarized antennas are commonly used in 
  	%base stations to provide diversity and improve capacity and coverage 
  	%\cite{Gabriel2016}, so 
  	%isolation by polarization is not a feasible option.

%While a handful works \cite{everett2016softnull} \rd{[more?]} have explored 
    Several questions nevertheless remain unanswered regarding the impact of 
    the array geometry on SI. For instance: \emph{which array geometries 
    achieve low SI and why?} An obvious solution to reducing SI is to space the 
    Tx and Rx arrays sufficiently far apart. However, this may be infeasible in 
    practice due to only a limited aperture or area being available for placing 
    the antenna elements. Placing all Tx and Rx antennas as far away as 
    possible from each other may also be undesirable due to the resulting 
    inefficient utilization of the available aperture/area, which a judicious \emph{sparse
    array} could leverage to improve the angular resolution \cite{sarangi2023superresolution} of both the 
    Tx and Rx arrays. The question, therefore, arises \emph{whether there are 
    array geometries that can simultaneously achieve high spatial resolution 
    and low SI?} Angular resolution is increasingly important in future 
    generation wireless systems operating at ever higher frequencies, where 
    line-of-sight channel conditions are prevalent and scattering is sparse. 
    Finally, for full-duplex massive MIMO systems employing large arrays, it 
    becomes important to understand how SI scales with the number of antennas 
    and aperture of the joint Tx-Rx array. Specifically, \emph{does SI 
    necessarily increase with a growing number of antennas?}

    \textbf{Contributions: }
    %based on spherical wave propagation, which is demonstrated to be representative of SI in antenna arrays with omnidirectional elements, such as ideal dipoles. We study% %To gain insight and obtain preliminary answers to the questions posed above, 
    This paper investigates how the array geometry impacts SI in full-duplex communications. We study a commonly considered idealized SI channel model and two elementary array configurations that are indicative of the best and worst-case SI within the class of linear arrays. We show that the SI channel matrix has a simple Toeplitz structure in case of both arrays. This paves the way for analytically characterizing the SI of a multitude of geometries, such as the nested array \cite{pal2010nested}, which we demonstrate can be adapted to the full-duplex setting to achieve both low SI and high angular resolution without ambiguities. We establish that the worst-case SI of the nested array need not increase with the number of antennas, but can actually decrease, provided the aperture also scales appropriately. This suggests that full-duplex massive MIMO systems with a large number of antennas may achieve negligible SI through array design alone.% constructing large full-duplex arrays with negligible SI, yet many antennas.% with the number of antennas. % as the array gets larger.% This structure allows analytically establishing a fundamental difference in the ... Furthermore, we show that combining allows  mitigation via antenna array design. Sparse arrays in literature.%constructed using these two prototypical arrays. We demonstrate how one such sparse array geometry

    \textbf{Notation: }
    Multiplication and addition of a set $\mathbb{X}$ and scalar $c\in\mathbb{R}$ are defined as $c\mathbb{X}=\{cx, x\in\mathbb{X}\}$ and $\mathbb{X}+c=\{x+c, x\in\mathbb{X}\}$, respectively. The largest (smallest) element of $\mathbb{X}\subset\mathbb{R}$ is denoted by $\max\mathbb{X}$ ($\min\mathbb{X}$). The set of the $N$ first non-negative integers is denoted by $\mathbb{U}_{N}\triangleq\{0,1,\ldots,N-1\}$.
    
%Renewed interest in \emph{full-duplex} (FD): Improving spectral efficiency in massive MIMO and emerging ISAC applications \cite{alexandropoulos2022fullduplex,barneto2021fullduplex}

%Judicious array design $\implies$ fewer valuable resources diverted from comms to mitigating SI \citep{everett2016softnull}

%Impact of array geometry of SI not well understood]
%\centering
%\begin{enumerate}[label=Q$_\arabic*$: ,leftmargin=2.5cm]
%	\item How does array geometry shape SI channel $\bm{H}_{\si}$?
%	\item To what extent can SI be mitigated via array design?
%\end{enumerate}
%
%\textbf{Contributions:}
%	\begin{enumerate}[label=\roman*),leftmargin=1.8cm]
%	\item Sparse arrays can \emph{enhance} spatial resolution while \emph{mitigating} self-interference (SI) in full-duplex systems% systems
%	\item SI \emph{need not} increase with \# of antennas for judiciously designed transmit-receive array configurations
%	\item \emph{Simple} SI models can yield \emph{insight} into the role of array geometry \& inform design of full-duplex systems% inform array design
%\end{enumerate}

    \section{Signal model}

    Consider a full-duplex base station (BS) with $N_{\tx}$ transmit (Tx) and $N_{\rx}$ receive (Rx) antennas. The BS simultaneously serves $D$ downlink (DL) and $U$ uplink (UL) users using the same spectral resources. Assuming a narrowband model, the received signal at the BS (UL model) can be written as
    \begin{align}
    \bm{y}=\bm{G}_{\ul}\bm{x}+\bm{H}_{\si}\bm{s}+\bm{n}, \label{eq:y}
    \end{align}
    where $\bm{x}\in\mathbb{C}^{U}$ is the symbol vector transmitted by the UL users, $\bm{s}\in\mathbb{C}^{N_{\tx}}$ the DL signal transmitted by the BS, and $\bm{n}\in\mathbb{C}^{N_{\rx}}$ denotes receiver noise. Furthermore, $\bm{G}_{\ul}\in\mathbb{C}^{N_{\rx}\times U}$ is the UL channel matrix, and $\bm{H}_{\si}\in\mathbb{C}^{N_{\rx}\times N_{\tx}}$ is the \emph{self-interference channel matrix}, of principle interest herein.% the principle focus of this paper.

% This will ultimately mitigating SI via antenna array design. Indeed, SI matrix % choosing the placement of the Tx and Rx antennas judiciously.% Simply subtracting $\bm{H}_{\si}\bm{s}$ from $\bm{y}$ is generally not possible, since $\bm{s}$ and $\bm{H}_{\si}$ may not be perfectly known at the transmitter due to transmitter noise and array calibration errors. Moreover, SI cannot typically be mitigated digitally at the receiver due to the risk of signal $\bm{H}_{\si}\bm{s}$ saturating the analog-to-digital converters, thereby masking the signal of interest $\bm{x}$. As a result, SI cancellation often entails either appropriate precoding at the transmitter [CITE] or analog cancellation at the receiver [CITE]. This paper considers a third, less explored approach to mitigating SI, namely \emph{antenna array design}. Indeed, SI matrix $\bm{H}_{\si}$ can be favorably shaped by choosing the placement of the Tx and Rx antennas judiciously.% including its spectral decomposition.%% A key challenge of full-duplex is the self-interference (SI) caused by transmitting $\bm{s}$ while simultaneously receiving% not only the reicever noise $\bm{n}$, but also
 \subsection{Self-interference model}  %matrix $\bm{H}_{\si}$
      %To gain insight into the impact of the Tx and Rx array geometries on SI,       
We consider a simple spherical wave SI model, where the $(n,m)$th entry of $\bm{H}_{\si}$ can be written as \cite{roberts2023lonestar} \footnote{Models similar to \eqref{eq:SI_model} have also been used for characterizing short-range MIMO communications channels \cite{jiang2005spherical} and mutual coupling \cite{liu2016supernested}.}%In addition to being regularly employed in full-duplex, m%Models similar to \eqref{eq:SI_model} have also been used
  \begin{align}
  %	[\bm{H}_{\si}]_{n,m}
  	{H_{\si}}_{n,m}= \rho\frac{\exp(j\pi\Delta_{n,m})}{\Delta_{n,m}}.\label{eq:SI_model}
  \end{align}
  Here, $\Delta_{n,m} \triangleq |d_{\rx}[n]-d_{\tx}[m]|$ denotes the distance between the $n$th Rx antenna $d_{\rx}[n]$ and $m$th Tx antenna $d_{\tx}[m] $ in units of $\lambda/2$, where $\lambda$ is the carrier wavelength. Consequently, $\bm{H}_{\si}=\bm{H}_{\si}(\mathbb{D}_{\rx},\mathbb{D}_{\tx}) $ is a function of the Tx and Rx array geometries $\mathbb{D}_{\tx}\triangleq\{d_{\tx}[m]\}_{m=1}^{N_{\tx}}$ and $\mathbb{D}_{\rx}\triangleq \{d_{\rx}[n]\}_{n=1}^{N_{\rx}}$, which, for simplicity, we assume to be one-dimensional and collinear. Moreover, $\rho=\rho(\mathbb{D}_{\rx},\mathbb{D}_{\tx})>0$ is a positive scaling factor, which may depend on the array geometry. \cref{eq:SI_model} is representative of SI in antenna arrays with omnidirectional elements, such as ideal dipoles, in an anechoic environment. In other cases,  deviation from \eqref{eq:SI_model} can be significant \cite{roberts2023spatial}.%is an appropriate model antennas with approximately omnidirectional radiation pattern, such as ideal dipoles.%Denote the normalized antenna positions in units of half carrier wavelength spacings by $\mathbb{D}_{\rx}=\{d_{\rx}[n]\}_{n=1}^{N_{\rx}}$ (Rx) and $\mathbb{D}_{\tx}=\{d_{\tx}[m]\}_{m=1}^{N_{\tx}}$ (Tx), where w.l.o.g. we assume $\mathbb{D}_{\rx},\mathbb{D}_{\tx}\subset \mathbb{N}$ are sets of non-negative integers.% denote the sets of normalized Tx and Rx antenna positions, respectively

\subsection{Objectives}
The goal of the BS is to recover (decode) UL signal $\bm{x}$. This task is complicated by the simultaneous transmission of DL signal $\bm{s}$, which couples back to the BS receiver. Indeed, SI signal $\bm{H}_{\si}\bm{s}$ can be much stronger than received signal $\bm{G}_{\ul}\bm{x}$. The objective of this paper is to \emph{gain an understanding of how matrix $\bm{H}_{\si}$ can be favorably shaped by antenna array design to mitigate SI}. More formally, an ideal $\bm{H}_{\si}$ would have a negligible worst-case SI with respect to the magnitude of the received signal of interest: $\|\bm{H}_{\si}\|_2\|\bm{s}\|_2\ll \|\bm{G}_{\ul}\bm{x}\|_2$. Alternatively, an approximately low rank $\bm{H}_{\si}$ may suffice such that most choices of $\bm{s}$ within some signal class of interest yield a small SI: $\|\bm{H}_{\si}\bm{s}\|_2\ll \|\bm{G}_{\ul}\bm{x}\|_2 \forall \bm{s}\in\mathcal{S}\subset \mathbb{C}^{N_{\tx}}$. For example, $\mathcal{S}$ could represent a Tx beamforming codebook designed for a millimeter-wave full-duplex communication system \cite{roberts2021millimeterwave,roberts2023lonestar}.\footnote{Similar conclusions hold for Rx beamforming (either in the digital or analog domain). Note that \eqref{eq:y} implicitly assumes a fully digital Rx architecture.} Achieving either of these objectives by array design alone reduces the need for additional SI mitigation, which could potentially save both hardware and computational resources.%\footnote{DL signal $\bm{s}$ may also be constrained to a subset of $\mathbb{C}^{N_{\tx}}$, such as in the case of an analog or hybrid Tx beamforming architecture. Similar conclusions hold for beamforming upon Rx (either in the digital or analog domain).} We note that %$\|\bm{H}_{\si}\bm{s}\|_2$ w.r.t. $\|\bm{G}_{\ul}\bm{x}\|_2$
%    \rd{[DL signal model for demonstrating angular resolution of Tx array? Mention effective Tx-Rx array (sum co-array) in case of JCS?]}

%    \subsubsection{Goal \labelcref{i:si}: Mitigating self-interference}

%    \subsubsection{Goal \labelcref{i:res}: Improving angular resolution}
%    Geometric UL channel model:
%    \begin{align*}
%        \bm{G}_{\ul}=\bm{A}_{\rx}(\bm{\theta})\diag(\bm{\alpha}),
%    \end{align*}
%    where $\bm{A}_{\rx}(\bm{\theta})\in\mathbb{C}^{N_{\rx}\times U}$ is the Rx array manifold matrix, $\bm{\theta}\in [-\frac{\pi}{2},\frac{\pi}{2}]^U$ and $\bm{\alpha}\in\mathbb{C}^U$ denote the channel coefficients.

    \section{Impact of array geometry on self-interference}\label{sec:structure}%Towards a theoretical understanding of the 
    %This section examines the impact that the Tx and Rx array geometries have on SI. 
    This section examines the structure of the SI matrix \eqref{eq:SI_model} in case of two elementary Tx-Rx array geometries denoted as the ``partitioned'' and ``interleaved'' arrays. This provides a preliminary understanding of how the joint Tx-Rx array geometry influences SI. For ease of exposition, we henceforth assume an equal number of Tx and Rx antennas $N=N_{\rx}=N_{\tx}$.

%     This paper considers the problem of designing the Tx and Rx array geometries to achieve two primary goals:
%    \begin{enumerate*}[label=\Roman*.,ref=\Roman*]
%    	\item Mitigating self-interference \label{i:si}
%    	\item Enhancing angular resolution.\label{i:res}
%    \end{enumerate*} 
    % attaining the ultimate objective of 
   %(prior to any additional SI cancellation, such as precoding at the tranasmitter or analog/hybrid beamforming at the receiver)

  \subsection{Structure of self-interference matrix}%: Case study%: Case of prototypical partitioned and interleaved array geometries
  
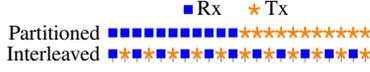
\begin{figure}
	\centering
	\newcommand{\Lmax}{21}
	\newcommand{\offs}{19}
	\newcommand{\msize}{1.5}
	\begin{tikzpicture} 
		\begin{axis}[width=5 cm ,height=2 cm,xmin=-0.2,xmax=\Lmax+0.2,ymin=-1.4,ymax=0.1,ytick style={draw=none},ytick={-1,0},yticklabels={Interleaved,Partitioned},xticklabel shift = 0 pt,xtick pos=bottom,axis line style={draw=none},legend style = {at={(0.5,1.03)},anchor=south,draw=none,fill=none},legend columns=2,xtick={0,1,...,\Lmax},xticklabels={}]%Nested,-2,
			\addplot[blue,only marks,mark=square*,mark size=\msize,y filter/.code={\pgfmathparse{\pgfmathresult*0}}] table[x=d,y=d]{Data/arrayRx_Partitioned_11_21.dat};
			\addplot[orange,only marks,mark=star,mark size=1.5*\msize,thick,y filter/.code={\pgfmathparse{\pgfmathresult*0}}] table[x=d,y=d]{Data/arrayTx_Partitioned_11_21.dat};
			\addlegendentry{Rx$\quad$}
			
			\addplot[blue,only marks,mark=square*,mark size=\msize,y filter/.code={\pgfmathparse{\pgfmathresult*0-1}}] table[x=d,y=d]{Data/arrayRx_Interleaved_11_21.dat};
			\addplot[orange,only marks,mark=star,mark size=1.5*\msize, thick,y filter/.code={\pgfmathparse{\pgfmathresult*0-1}}] table[x=d,y=d]{Data/arrayTx_Interleaved_11_21.dat};
			\addlegendentry{Tx}
			
%			\addplot[blue,only marks,mark=square*,mark size=\msize,y filter/.code={\pgfmathparse{\pgfmathresult*0-2}}] table[x=d,y=d]{Data/arrayRx_Nested_11_21.dat};
%			\addplot[orange,only marks,mark=star,mark size=1.5*\msize,thick,y filter/.code={\pgfmathparse{\pgfmathresult*0-2}}] table[x=d,y=d]{Data/arrayTx_Nested_11_21.dat};
		\end{axis}
	\end{tikzpicture}
	\caption{Partitioned and interleaved array geometries in \labelcref{eq:partitioned,eq:interleaved} with $N=11$ Tx/Rx antennas ($\delta_1=0,\delta_2=1$).}\label{fig:arrays_L_N}
	\end{figure}

 We define the partitioned and interleaved arrays as follows:
  \begin{itemize}[leftmargin=0cm]
  	\item[] \underline{\emph{Partitioned array}} (with parameter $\delta_1\in\mathbb{N}$):
  	\begin{align}
  		\begin{aligned}
  			\mathbb{D}_{\rx}&=\mathbb{P}_{\rx}\triangleq \mathbb{U}_N\\%\{0,1,\ldots,N-1\}\\
  			\mathbb{D}_{\tx}&=\mathbb{P}_{\tx}\triangleq \mathbb{P}_{\rx}+N+\delta_1,
  		\end{aligned}\label{eq:partitioned}
  	\end{align}
  \item[] \underline{\emph{Interleaved array}} (with parameter $\delta_2\in\mathbb{N}_+$):
  \begin{align}
  	\begin{aligned}
  	  	\mathbb{D}_{\rx}&=\mathbb{I}_{\rx}\triangleq 2\delta_2\mathbb{U}_N\\%\{0,1,\ldots,N-1\}\\
  	  	\mathbb{D}_{\tx}&=\mathbb{I}_{\tx}\triangleq \mathbb{I}_{\rx}+\delta_2.
	  \end{aligned}\label{eq:interleaved}
  \end{align}
    \end{itemize}
  In \eqref{eq:partitioned}, the joint Tx-Rx array is divided (partitioned) into two uniform linear arrays (ULAs), which are separated by $\delta_1$. Similarly, in \eqref{eq:interleaved}, the joint array geometry consists of alternating (interleaved) Tx and Rx antennas uniformly separated by $\delta_2$ (in units of $\lambda/2$). The partitioned and interleaved array configurations are depicted in \cref{fig:arrays_L_N} for $N=11$ Tx/Rx antennas, and parameters $\delta_1=0$, $\delta_2=1$. 
  
  The partitioned array is widely considered in the full-duplex literature, in 
  part due to its favorable SI compared to other linear arrays of equal 
  aperture and with the same number of antennas \cite{everett2016softnull}. 
  Hence, the partitioned array seems to be representative of the best-case 
  linear geometry in terms of SI.  In contrast, the interleaved array has been 
  observed to suffer from high SI \cite{everett2016softnull}, thus representing 
  a worst-case linear array geometry in terms of SI. While the interleaved 
  array is rarely employed in practice, it provides a useful model for 
  investigating the spectral properties of $\bm{H}_{\si}$, as we will see next. 
  Indeed, $\bm{H}_{\si}$ in \eqref{eq:SI_model} becomes a real-valued Toeplitz 
  matrix in case of both the partitioned and interleaved arrays. Specifically, 
  $\bm{H}_{\si}$ can be approximately low-rank for the partitioned array, 
  whereas this is the case for the interleaved array.% The nested array 
  %is deferred to future work.
  \vspace{-0.2cm}
	\subsubsection{Partitioned array} 
%	$\mathbb{D}_{\rx}=\{0,1,\ldots,N-1\}$ and $\mathbb{D}_{\tx}=\mathbb{D}_{\rx}+N+\delta_1$, where $\delta_1\in\mathbb{N}$. 
	
	In case of the partitioned array \eqref{eq:partitioned}, the distances between Tx-Rx antenna pairs form the following $N\times N$ Toeplitz matrix:
	\begin{align*}
		\bm{\Delta}(\mathbb{P}_{\rx},\mathbb{P}_{\tx})=
		\delta_1+
		\begin{bmatrix}
			N&N+1&\ldots&2N-1\\
			N-1&N&&2N-2\\
			\vdots& &\ddots &\vdots\\
			1&2 &\ldots&N
			\end{bmatrix}.
	\end{align*}
	Hence, $\bm{H}_{\si}$ in \eqref{eq:SI_model} is also a (real-valued) Toeplitz matrix:
 {\small
    \begin{align*}
    	\bm{H}_{\si}(\mathbb{P}_{\rx},\mathbb{P}_{\tx})
%    	&=
%    	e^{j\pi\delta_1}\begin{bmatrix}
%    		\frac{e^{j\pi N}}{N+\delta_1}&\frac{e^{j\pi(N+1)}}{N+1+\delta_1}&\ldots&\frac{e^{j\pi(2N-1)}}{2N-1+\delta_1}\\
%    		\frac{e^{j\pi(N-1)}}{N-1+\delta_1}&\frac{e^{j\pi N}}{N+\delta_1}&\ldots&\frac{e^{j\pi(2N-2)}}{2N-2+\delta_1}\\
%    		\vdots&&\ddots&\vdots\\
%    		\frac{e^{j\pi}}{1+\delta_1}&\frac{e^{j\pi 2}}{2+\delta_1}&\ldots&\frac{e^{j\pi N}}{N+\delta_1}
%    	\end{bmatrix}\\
    	&=
    	(-1)^{\delta_1}\rho_1\begin{bmatrix}
    		\frac{(-1)^N}{N+\delta_1}&\frac{(-1)^{N+1}}{N+1+\delta_1}&\ldots&\frac{-1}{2N-1+\delta_1}\\
    		\frac{(-1)^{N-1}}{N-1+\delta_1}&\frac{(-1)^{N}}{N+\delta_1}&\ldots&\frac{1}{2N-2+\delta_1}\\
    		\vdots&&\ddots&\vdots\\
    		\frac{-1}{1+\delta_1}&\frac{1}{2+\delta_1}&\ldots&\frac{(-1)^{N}}{N+\delta_1}
    	\end{bmatrix},
    \end{align*}
    }%
   for some $\rho_1>0$. The sign of the entries in $\bm{H}_{\si}$ alternates due to $\bm{\Delta}$ consisting of consecutive integers $n=1,2,\ldots, 2N-1$ and the phase term in \eqref{eq:SI_model} being of the form $e^{j\pi n}=(-1)^n$.

    \subsubsection{Interleaved array} 
    %	$\mathbb{D}_{\rx}=2\delta_2\{0,1,\ldots,N-1\}$ and $\mathbb{D}_{\tx}=\mathbb{D}_{\rx}+\delta_2$, where $\delta_2\in\mathbb{N}_+$. 
    
    In case of the interleaved array, distance matrix $\bm{\Delta}\in\mathbb{N}_+^{N\times N}$ is again Toeplitz due to the regular structure of the array:
    {\small
    \begin{align*}
    	\bm{\Delta}(\mathbb{I}_{\rx},\mathbb{I}_{\tx})=
    	\delta_2
    	\begin{bmatrix}
    		1&3&5&\ldots&2N-1\\
    		1&1&3& &2N-3\\
    		3&1&1& &2N-5\\
    		\vdots&         &          &\ddots &\vdots\\
    		2N-3   &2N-5&2N-7 &\ldots&1
    	\end{bmatrix}.
    \end{align*}
    }%
    Similarly, $\bm{H}_{\si}$ is also a real Toeplitz matrix (for any $\rho_2>0$):
    {\small
    \begin{align*}
    	\bm{H}_{\si}(\mathbb{I}_{\rx},\mathbb{I}_{\tx})
    	&=
    	\frac{(-1)^{\delta_2}\rho_2}{\delta_2}
    	\begin{bmatrix}
    		1&\frac{1}{3}&\frac{1}{5}&\ldots&\frac{1}{2N-1}\\
    		1&1&\frac{1}{3}&\ldots&\frac{1}{2N-3}\\
    		\frac{1}{3}&1&1&\ldots&\frac{1}{2N-5}\\
    		\vdots& & &\ddots&\vdots\\
    		\frac{1}{2N-3}&\frac{1}{2N-5}&\frac{1}{2N-7}&\ldots&1
    	\end{bmatrix}.
    \end{align*}
    }%
   All entries in $\bm{H}_{\si}$ have the same sign since $\bm{\Delta}$ consists of only odd or even integers depending on the value of $\delta_2\in\mathbb{N}_+$.

The structure of $\bm{H}_{\si}$ in case of the partitioned and interleaved arrays can be leveraged to gain insight into the fundamental dependence of SI on array geometry. Towards this goal, the next section inspects the scaling of the smallest and largest singular values of $\bm{H}_{\si}$ when $N=2$. This case study conveys the basic intuition behind why the partitioned array has a more favorable SI than the interleaved array. Extensions to arbitrary values of $N$ will be part of future work.

% To gain insight into the spectral properties of the SI matrices of the partitioned and interleaved arrays, we briefly consider the simple case $N=2$ next.% Extensions to arbitrary values of $N$ is left for future work.
    
\subsection{Self-interference in case of $N=2$ Tx \& Rx antennas}
%To gain insight into the spectral properties of the SI matrices of the partitioned and interleaved arrays, we briefly consider the simple case where $N=2$, i.e., 
When the Tx and Rx arrays both have two antennas each ($N=2$), the SI matrices of the partitioned and interleaved arrays, respectively, become
\begin{align*}
    \bm{H}_{\si}(\mathbb{D}_{\tx},\mathbb{D}_{\rx})
    =
    \begin{cases}
        (-1)^{\delta_1}\rho_1
	\begin{bmatrix}
		\frac{1}{2+\delta_1}&\frac{-1}{3+\delta_1}\\
		\frac{-1}{1+\delta_1}&\frac{1}{2+\delta_1}		
	\end{bmatrix},& \txt{if }
 \begin{cases} 
 \mathbb{D}_{\tx}=\mathbb{P}_{\tx}\\ \mathbb{D}_{\tx}=\mathbb{P}_{\rx}
 \end{cases}\\
    \\
	\frac{(-1)^{\delta_2}\rho_2}{\delta_2}\begin{bmatrix}
		1&\frac{1}{3}\\
		1&1
    \end{bmatrix}, & \txt{if } 
 \begin{cases}
 \mathbb{D}_{\tx}=\mathbb{I}_{\tx}\\
 \mathbb{D}_{\tx}=\mathbb{I}_{\rx}.
 \end{cases}
    \end{cases}
	% \bm{H}_{\si}(\mathbb{P}_{\tx},\mathbb{P}_{\rx})&=
	% (-1)^{\delta_1}c_1
	% \begin{bmatrix}
	% 	\frac{1}{2+\delta_1}&\frac{-1}{3+\delta_1}\\
	% 	\frac{-1}{1+\delta_1}&\frac{1}{2+\delta_1}		
	% \end{bmatrix},\\
	% \bm{H}_{\si}(\mathbb{I}_{\tx},\mathbb{I}_{\rx})&=
	% \frac{(-1)^{\delta_2}c_2}{\delta_2}\begin{bmatrix}
	% 	1&\frac{1}{3}\\
	% 	1&1
	% \end{bmatrix}.
\end{align*}
For large values of $\delta_1$, $\bm{H}_{\si}$ becomes approximately rank-1 in case of the partitioned array as $\bm{H}_{\si}(\mathbb{P}_{\tx},\mathbb{P}_{\rx})\approx (-1)^{\delta_1}\rho_1[\bm{h},-\bm{h}]$, where $\bm{h}=\frac{1}{\delta_1}[1,-1]^{\T}$. %More formally, when the partitioned and interleaved arrays have an equal aperture tending to infinity, the following result can be established.
%\begin{proposition}\label{thm:2x2}
%	Suppose $N=2$ and that the effective Tx-Rx array aperture $L\triangleq \max \mathbb{D}_{\tx}-\min \mathbb{D}_{\rx}$ is equal for both the partitioned  \eqref{eq:partitioned} and interleaved \eqref{eq:interleaved} arrays. Then
%	\begin{align*}
%		\lim_{L\to\infty}\frac{\|\bm{H}_{\si}(\mathbb{D}_{\tx},\mathbb{D}_{\rx})\|_2}{\|\bm{H}_{\si}(\mathbb{D}_{\tx},\mathbb{D}_{\rx})\|_{\F}}=		
%		\begin{cases}
%			1, & \txt{if }
%\begin{cases}
%    \mathbb{D}_{\tx}=\mathbb{P}_{\tx}\\
%    \mathbb{D}_{\rx}=\mathbb{P}_{\rx}
%\end{cases}\\%\txt{partitioned}\\
%			\sqrt{\frac{1}{2}+\frac{\sqrt{10}}{7}}<0.98,&
%   \txt{if }
%   \begin{cases}
%       \mathbb{D}_{\tx}=\mathbb{I}_{\tx}\\
%       \mathbb{D}_{\rx}=\mathbb{I}_{\rx}.
%   \end{cases}% \txt{interleaved}
%		\end{cases}
%	\end{align*}
%\end{proposition}
%\begin{proof}
%	Firstly, note that $L$ is the largest distance between any two antennas in the Tx-Rx array, since in \labelcref{eq:partitioned,eq:interleaved}, both $\max\mathbb{D}_{\rx} < \max \mathbb{D}_{\tx}$ and $\min\mathbb{D}_{\rx} < \min \mathbb{D}_{\tx}$.
%	Setting the aperture of the partitioned and interleaved arrays to be equal yields $2N-1+\delta_1=(2N-1)\delta_2\implies \delta_1=(2N-1)(\delta_2-1)$, where setting $N=2$ leads to $\delta_1=3(\delta_2-1)$.
% 
%    \rd{[TODO]}
%\end{proof}
%\cref{thm:2x2} shows that the  the SI matrix of the partitioned array becomes singular as the aperture increases. 
In contrast, the SI matrix of the interleaved array has full rank; in particular, the smallest and largest singular values are proportional to each other regardless of the value of $\delta_2$.\footnote{The singular values of $\bm{H}_{\si}(\mathbb{I}_{\tx},\mathbb{I}_{\rx})$ for $N=2$ can be verified to be $\sqrt{(14+4\sqrt{10})/9}\frac{\rho_2}{\delta_2}\approx 1.7\frac{\rho_2}{\delta_2}$ and $\sqrt{(14-4\sqrt{10})/9}\frac{\rho_2}{\delta_2}\approx 0.4\frac{\rho_2}{\delta_2}$.} This aligns with the intuition that when the separation between the Tx and Rx antennas grows large, all Rx antennas of the partitioned array observe the same \emph{single} plane wave originating from the direction of the Tx antennas, whereas for the interleaved array, all Rx antennas except one observe \emph{two} plane waves---one impinging from $-\pi/2$ and the other from $\pi/2$---since Tx antennas are located on both sides of these Rx antennas. This is also known as \emph{angular spread}---see \cite{everett2016softnull} and references therein.% \rd{[CITE angular spread literature. Transition into next section]}

    \section{Sparse arrays in full-duplex massive MIMO communications}
    
    This section demonstrates how judicious sparse array designs can yield both \emph{low SI} and \emph{high angular resolution} in full-duplex systems employing a large number of antennas. We explore the scaling of $\|\bm{H}_{\si}\|_2$, which controls the worst-case SI, as a function of the number of antennas and array aperture. Contrary to \cref{sec:structure}, the entries of $\bm{H}_{\si}$ are obtained from the S-parameters---simulated via the MATLAB antenna toolbox \cite{matlab2022antenna}---of the $N_{\rx}\times N_{\tx}$-port system corresponding to the joint Tx-Rx antenna array, assumed to consist of ideal strip dipole antennas (of length $\lambda/2$ and width $\lambda/100$) in an anechoic environment. Future work will investigate the scaling of SI using \eqref{eq:SI_model}, which may also require modeling scalar $\rho$.% that can yield further insight into the .%, provided an accurate.%Future work will explore the analytical SI model \eqref{eq:SI_model}, which can provide more insight into the scaling of SI, provided an accurate model for scaling factor $c$ is available for the considered array geometries.
   % The analytical SI model \eqref{eq:SI_model} can provide insight into the scaling of SI, provided an accurate model for scaling factor $c$ is available for the considered array geometries. %Indeed, the simple structure of $\bm{H}_{\si}$  in \eqref{eq:SI_model} permits writing the Frobenius norm $\|\bm{H}_{\si}\|_{\F}$ as a function of \emph{generalized harmonic numbers} [CITE] of the form $\sum_{\ell =1}^a 1/\ell^q$ for some positive integers $a,q\in\mathbb{N}_+$. Hence, the SI (or a proxy thereof, given by $\|\bm{H}_{\si}\|_{\F}$) of the partitioned, interleaved or nested array could be obtained in closed-form as a function of the number of antennas $N$ and Tx-Rx array separation $\delta$. 
   % We defer this analysis to future work.%Instead of \eqref{eq:SI_model}, t%Can we gain insight into SI scaling laws and dependence on geometry w/o costly electromagnetic simulations?% Specifically, we focus on the nested array \cite{pal2010nested}, whose simple structure\ldots \rd{[TODO]}
    %can be determined more precisely via measurement or numerical electromagnetic solvers. This paper will use the latter approach as a baseline to complement and verify \eqref{eq:SI_model}. In this case, 

    \subsection{Nested full-duplex array: Low self-interference and high resolution without ambiguities}
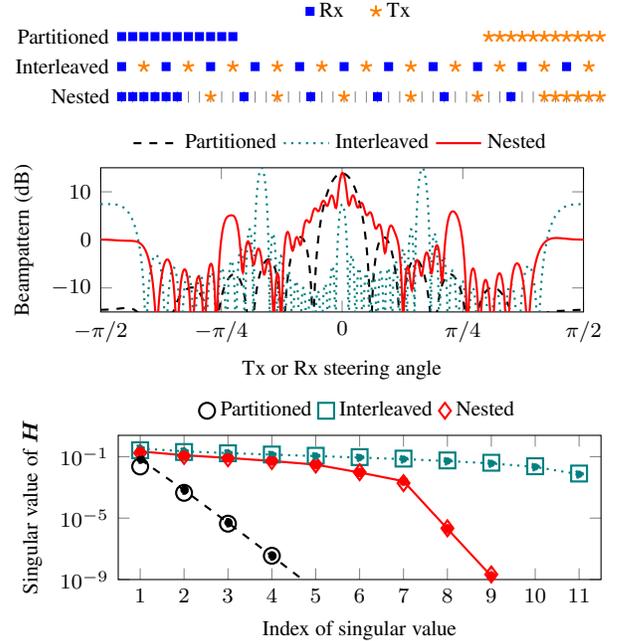
\begin{figure}
    	\centering
    	\newcommand{\figwidth}{8}
    	\newcommand{\Lmax}{43}
    	\newcommand{\offs}{19}
    	\newcommand{\msize}{1.5}
     %   \subfloat[Array geometries]{    
    	\begin{tikzpicture} 
    		\begin{axis}[width=\figwidth cm ,height=2.5cm,xmin=-0.2,xmax=\Lmax+0.2,ymin=-2.2,ymax=0.1,ytick style={draw=none},ytick={-2,-1,0},yticklabels={Nested,Interleaved,Partitioned},xticklabel shift = 0 pt,xtick pos=bottom,axis line style={draw=none},legend style = {at={(0.5,1.03)},anchor=south,draw=none,fill=none},legend columns=2,xtick={0,1,...,\Lmax},xticklabels={}]%,xtick={0,2,...,\Lmax}
    			\addplot[blue,only marks,mark=square*,mark size=\msize,y filter/.code={\pgfmathparse{\pgfmathresult*0}}] table[x=d,y=d]{Data/arrayRx_Partitioned_11_43.dat};
    			\addplot[orange,only marks,mark=star,mark size=1.5*\msize,thick,y filter/.code={\pgfmathparse{\pgfmathresult*0}}] table[x=d,y=d]{Data/arrayTx_Partitioned_11_43.dat};
    			\addlegendentry{Rx$\quad$}
    			
    			\addplot[blue,only marks,mark=square*,mark size=\msize,y filter/.code={\pgfmathparse{\pgfmathresult*0-1}}] table[x=d,y=d]{Data/arrayRx_Interleaved_11_42.dat};
    			\addplot[orange,only marks,mark=star,mark size=1.5*\msize, thick,y filter/.code={\pgfmathparse{\pgfmathresult*0-1}}] table[x=d,y=d]{Data/arrayTx_Interleaved_11_42.dat};
    			\addlegendentry{Tx}
    			
    			\addplot[blue,only marks,mark=square*,mark size=\msize,y filter/.code={\pgfmathparse{\pgfmathresult*0-2}}] table[x=d,y=d]{Data/arrayRx_Nested_11_43.dat};
    			\addplot[orange,only marks,mark=star,mark size=1.5*\msize, thick,y filter/.code={\pgfmathparse{\pgfmathresult*0-2}}] table[x=d,y=d]{Data/arrayTx_Nested_11_43.dat};
    		\end{axis}
    	\end{tikzpicture}
     %\label{fig:arrays_L_N2}}
    	\newcommand{\figheight}{3.5}
    % \subfloat[Beampattern]{ 
    	\begin{tikzpicture}
    		\begin{axis}[width=8 cm,height=\figheight cm,ymin=0,ylabel={Beampattern (dB)},xmin=-pi/2,xmax=pi/2,xlabel={Tx or Rx steering angle},xtick={-pi/2,-pi/4,0,pi/4,pi/2},xticklabels={$-\pi/2$,$-\pi/4$,$0$,$\pi/4$,$\pi/2$},ymin=-15,ymax=15,title style={yshift=0 pt},xticklabel shift = 0 pt,xlabel shift = {0 pt},yticklabel shift=0pt,ylabel shift = 0 pt,legend style = {at={(0.5,1.03)},anchor=south,draw=none,fill=none},legend columns=3]%ymode=log,ymin=0.001,ymax=1.2
    			\addplot[black, thick,dashed] table[x=theta,y=B]{Data/beampattern_Partitioned_11_11_43.dat};
    			\addlegendentry{Partitioned}
    			\addplot[teal,thick, dotted] table[x=theta,y=B]{Data/beampattern_Interleaved_11_11_42.dat};
    			\addlegendentry{Interleaved}
    			\addplot[red, thick] table[x=theta,y=B]{Data/beampattern_Nested_11_11_43.dat};
    			\addlegendentry{Nested}
    		\end{axis}
    		% \begin{axis}[
    			% 	width=30 cm,height=10cm,
    			%  %hide axis,
    			%  %scale only axis,
    			%  legend style = {draw=none,fill=none},legend columns=3
    			%   ]
    			%      \addplot[black,very thick,draw=none] table[x=theta,y=B]{Data/beampattern_Partitioned_11_11_43.dat};
    			%      \addlegendentry{Partitioned}
    			%      \addplot[gray,very thick, dashed] table[x=theta,y=B]{Data/beampattern_Interleaved_11_11_42.dat};
    			%      \addlegendentry{Interleaved}
    			%  \end{axis}
    	\end{tikzpicture}
%     \label{fig:arrays_L_N2_bp}}\\
 %    \subfloat[Singular values]{ 
     \newcommand{\const}{2e-1}
      \newcommand{\opac}{0.2}
      \newcommand{\msizea}{0.7}
    	\begin{tikzpicture} %, $i$
    		\begin{axis}[width=8 cm,height= \figheight cm,ylabel={Singular value of $\bm{H}_{\si}$},xlabel= {Index of singular value},xmin=0.5,xmax=11.5,xlabel shift = {0 pt},xtick={1,2,...,11},ymin=1e-9, ymode=log,legend style = {at={(0.5,1.03)},anchor=south,draw=none,fill=none},legend columns=3]% yticklabel pos=right,ylabel={},clip=false,
    			\addplot+[only marks,black, mark = o,mark options={scale=\msize}, thick] table[x=i,y =sigma]{Data/svd_Partitioned_Nt_11_Nr_11_L_43.dat};
    			\addlegendentry{Partitioned}
    			\addplot+[only marks,teal, thick,dashed, mark = square,mark options={scale=\msize,solid}] table[x=i,y =sigma]{Data/svd_Interleaved_Nt_11_Nr_11_L_42.dat};%
    			\addlegendentry{Interleaved}
    			\addplot+[only marks,red, thick,dashed, mark = diamond,mark options={scale=\msize,solid}] table[x=i,y =sigma]{Data/svd_Nested_Nt_11_Nr_11_L_43.dat};%ycomb
    			\addlegendentry{Nested}
           \addplot+    [mark=*,mark options={scale=\msizea,draw opacity=0},black,dashed,thick,draw opacity=\opac,fill opacity=\opac] table[x=i,y expr =\thisrow{sigma}*\const]{Data/svd_model_Partitioned_Nt_11_Nr_11_L_43.dat};%, dashed
			%\addlegendentry{Part. ($\bm{\bar{H}}_{\si}$)$\ $ }
			\addplot+[mark=*,mark options={scale=\msizea,draw opacity=0},fill opacity=\opac,teal,dotted, thick,draw opacity=\opac] table[x=i,y expr=\thisrow{sigma}*\const]{Data/svd_model_Interleaved_Nt_11_Nr_11_L_42.dat};%
			%\addlegendentry{Intrl. ($\bm{\bar{H}}_{\si}$)$\ $}
			\addplot+[mark=*,mark options={scale=\msizea,draw opacity=0},red, thick,solid,draw opacity=\opac,fill opacity=\opac] table[x=i,y expr=\thisrow{sigma}*\const]{Data/svd_model_Nested_Nt_11_Nr_11_L_43.dat};%ycomb
			%\addlegendentry{Nst. ($\bm{\bar{H}}_{\si}$)}
    		\end{axis}%
    	\end{tikzpicture}
%     \label{fig:arrays_L_N2_sv}}
     \caption{The nested full-duplex array leverages the full aperture of the joint Tx-Rx array (top) to achieve a narrow Tx/Rx main lobe width (mid) and SI between that of the partitioned and interleaved arrays (bottom). The SI of the simulated dipole array is well approximated by the spherical wave model \eqref{eq:SI_model} (bottom: solid marks vs. transparent lines) ($N=11$, $\delta_1=23$, $\delta_2=2$, $M_1=6$ $M_2=5$, $\delta_3=3$).}\label{fig:arrays_L_N2}%The bottom plot also demonstrates the agreement between the spherical wave model in \eqref{eq:SI_model} (transparent lines) and the simulated dipole array (solid marks)
    \end{figure}

    Increasing the array aperture for a fixed number of (Tx and Rx) antennas 
    has two key advantages. Firstly, it can decrease SI. Secondly, the 
    increased aperture enables enhancing angular resolution. However, 
    leveraging both reduced SI and increased aperture requires judiciously 
    designing the joint Tx-Rx array geometry. Indeed, both the partitioned and 
    interleaved arrays fall short in this aspect: in case of the partitioned 
    %RW PItäiskö tässä olla P_rx jne.
    array, both the Rx and Tx array apertures ($\max 
    \mathbb{D}_{\rx}-\min\mathbb{D}_{\rx}$ and $\max 
    \mathbb{D}_{\tx}-\min\mathbb{D}_{\tx}$) are fixed regardless of the 
    effective aperture of the joint Tx-Rx array $L\triangleq\max 
    \mathbb{D}_{\tx}-\min\mathbb{D}_{\rx}$; in case of the interleaved array, 
    the uniform sub-sampling of the Tx and Rx arrays gives rise to grating 
    lobes, which in turn lead to angular ambiguities. %This is demonstrated in 
    %\cref{fig:arrays_L_N2} for $N=11$ and array parameters 
    %$\delta_1=23,\delta_2=2$.
    The limited resolution of the partitioned array and the grating lobes of the interleaved array can be overcome by employing a sparse array geometry, such as the following variation of the widely-known \emph{nested array} \cite{pal2010nested}:    \begin{itemize}[leftmargin=0cm]
        \item[] \underline{\emph{Nested (full-duplex) array}} (with parameters $M_1,\!M_2,\!\delta_3\!\in\!\mathbb{N}_+$)
    \begin{align}
    \begin{aligned}
        \mathbb{D}_{\rx}&=
        % \mathbb{S}_{\rx}\triangleq \{0,1,\ldots,M_1-1\}\cup (2M_1+M_1\{0,1,\ldots,M_2-1\})\\
        \mathbb{S}_{\rx}\triangleq \mathbb{U}_{M_1}\cup (2\delta_3(\mathbb{U}_{M_2}+1)+M_1-1)\\
        \mathbb{D}_{\tx}&=\mathbb{S}_{\tx}\triangleq \max \mathbb{S}_{\rx} -\mathbb{S}_{\rx}+M_1-1+\delta_3.
        \end{aligned}\label{eq:nested}
    \end{align}
     \end{itemize}
   Here, $M_1,M_2$ are positive integers satisfying $N=M_1+M_2$ for a given $N$, and $\delta_3$ controls the inter-sensor spacing of the sparse ULA. The nested array can be seen as a mixture of the partitioned and interleaved arrays, as illustrated in \cref{fig:arrays_L_N2} (top) for $N=11$ and parameter values $\delta_1=23$, $\delta_2=2$, $M_1=6$, $M_2=5$, $\delta_3=3$. The nested array leverages the full aperture $L$ of the joint Tx-Rx array to achieve a narrow (Tx or Rx) main beam without introducing angular ambiguities via high side lobes (\cref{fig:arrays_L_N2}, center). It also strikes a balance between the SI of the partitioned and interleaved arrays, as evidenced by the singular values of $\bm{H}_{\si}$ shown in the bottom plot of \cref{fig:arrays_L_N2}. The transparent lines correspond to the spherical wave model in \eqref{eq:SI_model} with $\rho=0.2$. The agreement with the electromagnetic solver (solid markers) is apparent, which further validates the physical relevance of \eqref{eq:SI_model}. 
   
  Finally, we note that when $L\propto N^2$, the nested full-duplex array can also be verified to yield on the order of $N^2$ contiguous \emph{virtual} Tx-Rx antennas in its so-called \emph{sum co-array}. The sum co-array \cite{hoctor1990theunifying} is a virtual array arising in active sensing systems with co-located transmitters and receivers. The large contiguous co-array guarantees the identifiability of $\mathcal{O}(N^2)$ scatterers/targets, which is desirable when the BS also performs sensing, as is the case in dual-function ISAC systems.    
    \vspace{-1cm}

\subsection{Scaling of self-interference with $N$}

%Next, we compare the SI of the partitioned, interleaved, and nested array as a function of the number of antennas $N$. The regime of large $N$ is of interest in massive MIMO and other future wireless systems requiring high spatial resolution.% Caution should nevertheless be exercised, as scaling factor $c$ in \eqref{eq:SI_model} may also be a function of $N$ and $\delta$. Consequently, we leave a detailed investigation of this for future work.

%Note that the simple structure of $\bm{H}_{\si}$ permits writing the Frobenius norm $\|\bm{H}_{\si}\|_{\F}$ as a function of \emph{generalized harmonic numbers} [CITE] of the form $\sum_{\ell =1}^a 1/\ell^q$ for some positive integers $a,q\in\mathbb{N}_+$. Hence, the SI of the partitioned array (or a proxy thereof, given by $\|\bm{H}_{\si}\|_{\F}$) could be obtained in closed-form as a function of the number of antennas $N$ and Tx-Rx array separation $\delta$. Caution should nevertheless be exercised, as scaling factor $c$ in \eqref{eq:SI_model} may also be a function of $N$ and $\delta$. Consequently, we leave a detailed investigation of this for future work.

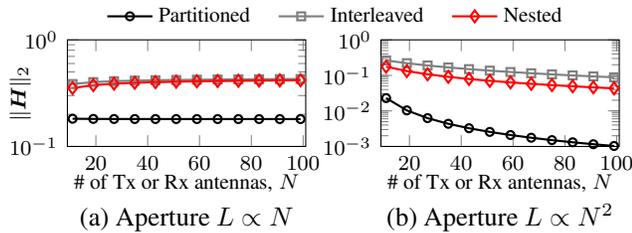
\begin{figure}
  \newcommand{\pnorm}{specn}
\newcommand{\msize}{1.5}
\begin{tikzpicture}
	\begin{groupplot}[
		group style={
			group size=2 by 1,
			%x descriptions at=edge bottom,
			horizontal sep=1cm,
			vertical sep=0cm,
		},
		width=4.75 cm,height=3.0 cm,ymin=0,ylabel={$\|\bm{H}_{\si}\|_2$},xlabel={\# of Tx or Rx antennas, $N$},xticklabel shift = 0 pt,yticklabel shift=0pt,ymode=log,xmin=9,xmax=101,title style={yshift=-82 pt},xlabel shift = {-5 pt},ylabel shift = {-15 pt}
		]%
		\nextgroupplot[%
		,legend to name=grouplegend
		,mark=none,legend style = {draw=none,fill=none},legend columns=3,ymin=0.1,ymax=1,title={\begin{enumerate*}[label=(\alph*)]
				\item Aperture $L\propto N$\label{fig:LproptoN}
			\end{enumerate*}}%, title={(a) Aperture $L\propto N$},title style={at={(0.5,-0.7)}}
		]
		\addplot[black, thick,mark=o,mark size=\msize] table[x=Nt,y=\pnorm]{Data/SInorm_Partitioned_Lmax_197.dat};
		\addlegendentry{Partitioned$\quad$}
		\addplot[gray,thick,mark=square,mark size=\msize] table[x=Nt,y=\pnorm]{Data/SInorm_Interleaved_Lmax_197.dat};
		\addlegendentry{Interleaved$\quad$}
		\addplot[red,thick,mark=diamond,mark size=1.5*\msize] table[x=Nt,y=\pnorm]{Data/SInorm_Nested_Lmax_197.dat};
		\addlegendentry{Nested}
		%\node[] at (axis cs: 55,8e-2) {\footnotesize{Aperture $L\propto N$}};
		\nextgroupplot[ylabel={},ymin=1e-3,ymax=1e0,title={\begin{enumerate*}[label=(\alph*)]
				\setcounter{enumi}{1}
				\item Aperture $L\propto N^2$\label{fig:LproptoN2}
			\end{enumerate*}}]
		\addplot[black,thick,mark=o,mark size=\msize] table[x=Nt,y=\pnorm]{Data/SInorm_Partitioned_Lmax_2573.dat};
		\addplot[gray,thick,mark=square,mark size=\msize] table[x=Nt,y=\pnorm]{Data/SInorm_Interleaved_Lmax_2573.dat};
		\addplot[red,thick,mark=diamond,mark size=1.5*\msize] table[x=Nt,y=\pnorm]{Data/SInorm_Nested_Lmax_2573.dat};
		%\node[] at (axis cs: 55,55e-3) {\footnotesize{Aperture $L\propto N^2$}};
	\end{groupplot}
	\node at (group c1r1.north) [anchor=north, yshift=.7cm, xshift=2cm] {\ref{grouplegend}};
\end{tikzpicture}\vspace{-0.4cm}
\caption{The largest singular value of $\bm{H}$ can decrease with the number of Tx/Rx antennas $N$ if the aperture $L$ grows suitably with $N$ and the array geometry is such that the spacing between each Tx-Rx antenna pair also increases with $N$.}\label{fig:si_scaling}%, provided the array geometry is such that the spacing between each Tx-Rx antenna pair also increases with $N$% For each of the considered \labelcref{fig:LproptoN} %Largest singular value of SI matrix as a function of the number of Tx/Rx antennas $N$. Self-interferenc
\end{figure}

\cref{fig:si_scaling} shows the largest singular value of $\bm{H}_{\si}$ as a function of the number of Tx/Rx antennas $N$ for two different aperture scaling laws, $L\propto N$ and $L\propto N^2$. The regime of large $N$ and $L$ is of interest in future wireless systems requiring high spatial resolution.\footnote{When $L\propto N^2$, the main lobe width of the Tx/Rx array beampattern is proportional to $1/N^2$ in case of the nested and interleaved arrays, whereas it is only proportional to $1/N$ in case of the partitioned array.} When $L\propto N$, $\|\bm{H}_{\si}\|_2$ only weakly depends on $N$. However, as $L\propto N^2$, $\|\bm{H}_{\si}\|_2$ decreases with $N$ since the separation between \emph{any} Tx-Rx antenna pair also increases in case of the three considered array geometries. This suggests that massive MIMO systems can actually mitigate SI more effectively when the number of antennas increases by virtue of judicious array design alone. %Interestingly, this suggests the existence of aperture scaling laws $L\propto N^\alpha$, where $1<\alpha<2$, for which $\|\bm{H}_{\si}\|_2$ is constant as a function of $N$. The value of $\alpha$ for which this occurs generally depends on the array configuration. Analytically characterizing this threshold, for example using \eqref{eq:SI_model} given a model for scaling factor $c$, is part of ongoing work.%

%Regarding resolution, it is well-known that the main lobe width (MLW) of the array beampattern is inversely proportional to the aperture []. Hence, the MLW of the nested array is proportional to $1/N^2$ (for both the Tx and Rx arrays), whereas for the partitioned array it is only proportional to $1/N$ []. While the MLW of the interleaved array is comparable to that of the nested array, the interleaved array suffers from severe grating lobes which decreases its unambiguous field-of-view (cf. \cref{fig:arrays_L_N2}).

\vspace{-.55cm}

%\rd{[Example with numbers here to illustrate SI and address question of absolute acceptable SI level?]}

    \section{Conclusions}
    This paper investigated the impact of the array geometry on self-interference (SI) in wireless full-duplex systems. We showed that a simple spherical-wave model can provide insight into the degree of SI caused by different array geometries. Specifically, the SI matrix assumes a simple Toeplitz structure in case of certain elementary array configurations consisting of ULAs (including uniformly subsampled ones). Hence, it becomes possible to analytically characterize the spectral properties of the SI matrix, which can provide a preliminary understanding of the SI experienced by different array geometries as a function of aperture and the number of antennas. We demonstrated that judicious sparse array designs, such as the nested array, are able to leverage an increased joint Tx-Rx aperture for achieving both low self-interference and high angular resolution (upon both Tx and Rx). Simulations showed that SI can decrease with an increasing number of Tx/Rx antennas for such sparse arrays. Future work will examine SI in case of both one and two-dimensional array geometries at depth.% linear arrays, as well as extending this analysis to (sparse) planar arrays which are often employed in practice.%revealing that SI need not grow---but can actually decrease---with the number of Tx/Rx antennas $N$, provided the spacing between each Tx-Rx antenna pair increases as a suitable (array-dependent) function of $N$. %$N$, provided the spacing between each Tx-Rx antenna pair increases as a suitable (array-dependent) function of $N$
    \vspace{-0.4cm}

	\bibliographystyle{IEEEtran}
	% Onkohan tämä hyväksyttävää
	\let\oldthebibliography\thebibliography
	\let\endoldthebibliography\endthebibliography
	\renewenvironment{thebibliography}[1]{
	  \begin{oldthebibliography}{#1}
	    \setlength{\itemsep}{0em}
	    \setlength{\parskip}{0em}
	}
	{
	  \end{oldthebibliography}
	}
  \bibliography{IEEEabrv,references}%../
	
	% that's all folks
\end{document}